\documentclass{optica-article}

\journal{optcon}

\articletype{Research Article}

\usepackage{verbatim}

\usepackage{lineno} 
\usepackage{amsmath}           
\usepackage{etoolbox}          

\newcommand*\linenomathpatch[1]{%
  \cspreto{#1}{\linenomath}%
  \cspreto{#1*}{\linenomath}%
  \csappto{end#1}{\endlinenomath}%
  \csappto{end#1*}{\endlinenomath}%
}
\newcommand*\linenomathpatchAMS[1]{%
  \cspreto{#1}{\linenomathAMS}%
  \cspreto{#1*}{\linenomathAMS}%
  \csappto{end#1}{\endlinenomath}%
  \csappto{end#1*}{\endlinenomath}%
}

\expandafter\ifx\linenomath\linenomathWithnumbers
  \let\linenomathAMS\linenomathWithnumbers
  \patchcmd\linenomathAMS{\advance\postdisplaypenalty\linenopenalty}{}{}{}
\else
  \let\linenomathAMS\linenomathNonumbers
\fi

\linenomathpatch{equation}
\linenomathpatchAMS{gather}
\linenomathpatchAMS{multline}
\linenomathpatchAMS{align}
\linenomathpatchAMS{alignat}
\linenomathpatchAMS{flalign}

\makeatletter
\patchcmd{\mmeasure@}{\measuring@true}{
  \measuring@true
  \ifnum-\linenopenaltypar>\interdisplaylinepenalty
    \advance\interdisplaylinepenalty-\linenopenalty
  \fi
  }{}{}
\makeatother

\newcommand{\be}[0]{\begin{equation}}
\newcommand{\ee}[0]{\end{equation}}
\newcommand{\bea}[0]{\begin{eqnarray}}
\newcommand{\eea}[0]{\end{eqnarray}}

\newcommand{\ud}[0]{{\rm d}}
\newcommand{\ii}[0]{{\rm i}}

\newcommand{\Tr}{\text{Tr}}

\begin{document}

\title{Off-axis Aberrations Improve the Resolution Limits of Incoherent Imaging}

\author{Kevin Liang\authormark{1,2,*}}

\address{\authormark{1}Physics Department, Adelphi University, Garden City, NY 11530, USA\\
\authormark{2}The Institute of Optics, University of Rochester, Rochester NY 14627, USA\\
}

\email{\authormark{*}kliang@adelphi.edu} 



\begin{abstract*}
The presence of off-axis tilt and Petzval curvature, two of the lowest-order off-axis Seidel aberrations, is shown to improve the Fisher information of two-point separation estimation in an incoherent imaging system compared to an aberration-free system. Our results show that the practical localization advantages of modal imaging techniques within the field of quantum-inspired superresolution can be achieved with direct imaging measurement schemes alone.
\end{abstract*}

\section{Introduction}

In well-corrected imaging systems, the propagation from the object to image planes is often modeled with shift invariance \cite{Goodman}. This indicates that the profile of the imaging system's point spread function (PSF) is independent of object impulse's location. Recent works have recasted the associated Rayleigh's criterion, which states that the separation of two nearby point sources cannot be precisely estimated, for such systems in terms of the classical Fisher information (CFI) \cite{Helmstrom}. The CFI informs on the precision one expects in extracting an unknown parameter with a given measurement scheme. In this context, Rayleigh's criterion states that the CFI for the separation of two point sources, with direct intensity (DI) measurements, vanishes as the separation between the point sources approaches zero. 

A shift in the well-established understanding of the Rayleigh's criterion was provided by Tsang and Nair \cite{Tsang2016}. They showed that the quantum Fisher information (QFI), an upperbound for the CFIs over all possible measurement schemes, for the separation of two equally bright incoherent point sources remains non-zero in the sub-Rayleigh regime. This result, along with the introduction of modal imaging techniques like spatial mode demultiplexing (SPADE), a measurement scheme that saturates the QFI, is responsible for the wide net of recent research efforts that sought to extend the promise of this so-called quantum-inspired superresolution \cite{Tsang2018,Tsang2019,Tsang:19,stanislaw,ultimateTiming,ultimate2,Zhou2019}. Such extensions include theoretical work in which strides have been made regarding the generalization of the object's spatial and coherence details \cite{Larson2018,Liang2021,Liang2021B,Liang2022rev,Bisketzi_2019,stanislaw,Hradil2019}. Regarding experimental considerations, the advantages of SPADE-type measurements have been tested and confirmed for a variety of objects \cite{Wadood2021,ultimateTiming}.

In this work, calculations are provided for the CFI and QFI for the separation between two equally bright incoherent point sources in imaging systems which possess off-axis/field-dependent aberrations. The presence of these aberrations causes imaging systems to be shift-variant; the analysis of such systems have been thus far neglected in the context of quantum-inspired superresolution. Remarkably, we show that the inclusion of simple low-order off-axis Seidel aberrations can give rise to more precise estimation of source separation when compared to an aberration-free system by improving on the CFI and QFI. In particular, we provide the analysis for off-axis tilt (OAT) and Petzval curvature. Both are shown to improve the CFI for both DI and modal imaging measure schemes, with the former providing a global improvement and the latter giving rise to improvements in the sub-Rayleigh regime. To derive these results, a general framework for shift-variant imaging systems is introduced in Section~\ref{theory}, from which a specialized treatment for two equally bright incoherent point sources is considered. Our results, given in Section~\ref{section:res}, provide context to the nascent field of quantum-inspired superresolution and show that much improvement can be made through the intentional usage of off-axis aberrations.

\section{Theoretical Background} \label{theory}
\subsection{Shift variant imaging systems}
\begin{figure}
    \centering
    \includegraphics[scale = 0.75]{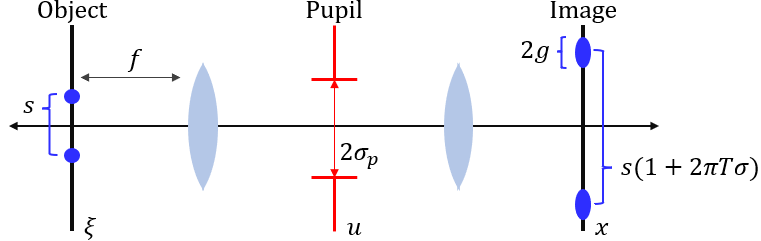}
    \caption{A schematic of a $4f$ imaging system with $\xi,u,$ and $x$ as the spatial coordinates for the object, pupil, and image planes, respectively. An illustration of two point sources separated by $s$ is shown along with their image distributions in the presence of OAT and Petzval curvature of strengths $T$ and $P$, respectively, which are defined in Eq.~(\ref{strparams}). The individual PSFs have a half-width of $g$ (which depends on $P$), defined in Eq.~(\ref{PSFwidth}), and their separation is magnified by a factor of $1 + 2\pi T \sigma$. The pupil is modeled as a Gaussian with half-width $\sigma_p$.}
    \label{4fsystem}
\end{figure}

In a shift-invariant imaging system, the optical field at the image plane may be obtained from the optical field at the object plane via a convolution with the system's PSF, $\psi$ \cite{Goodman}. In the presence of off-axis aberrations, however, the relation between object field, $U_o$, and image field, $U_i$, is given by a more general integral transformation. For simplicity, we assume one transverse spatial coordinate for which the aforementioned transformation is given by
\begin{align}
    U_i(x) = \int_{-\infty}^\infty U_o(\xi) \psi(x,\xi)\, \ud \xi, \label{convreplace}
\end{align}
where $\xi$ and $x$ are the object and image plane spatial coordinates, respectively. The departure from a shift-invariant imaging system is capture by $\psi(x,\xi)$, which indicates that the system PSF may have a dependence on both $\xi$ and $x$ that is not confined to their difference $x-\xi$. The form for $\psi(x,\xi)$ may be obtained by considering the propagation of a quasi-monochromatic point source with wavenumber $k = 2\pi/ \lambda$, modeled as a Dirac delta impulse, located at $\xi$, through the imaging system. This impulse is mapped onto a linear phase factor at the pupil plane. There, the linear phase factor is modulated by a Gaussian pupil function of characteristic width $\sigma_p$ and a phase aberration function, $\Delta W$. The choice of a Gaussian pupil function, commonly made in related works, is used for mathematical convenience \cite{Tsang2016,Liang2021,Larson2018}. An inverse Fourier transformation is then performed on the pupil-plane field to give
\begin{align}
    \psi(x,\xi) = U_0 \int_{-\infty}^\infty \exp\left( - \ii k \frac{\xi u}{ f} \right)  \exp\left( - \frac{u^2}{4 \sigma_p^2} \right)  \exp \left[ -\ii k \Delta W(u,\xi) \right] \exp \left(\ii k  \frac{ux}{ f} \right)\, \ud u, \label{SVPSF}
\end{align}
where $U_0$ is a constant with dimensions of optical field that ensures the intensity-normalization of $\psi$. Furthermore, $u$ is the pupil-plane spatial coordinate, and $f$ is the focal length of the lenses in a $4f$ imaging system (although the results presented in this work are valid for arbitrary other imaging configurations); the Fourier optical details regarding Eq.~(\ref{SVPSF}) is described further in \textcolor{blue}{Supplement 1}. A schematic of the imaging system is shown in Fig.~\ref{4fsystem}. It will be useful to define $\sigma = f\lambda /(4\pi \sigma_p)$ as the characteristic width of the diffraction-limited [$\Delta W(u,\xi) = 0$] PSF. Notice, as is true for the case of a shift-invariant imaging system, the system PSF would explicitly be a function of the difference in coordinates $x - \xi$ if $\Delta W$ were independent of the object location $\xi$. Such on-axis/field-independent aberrations include well-known Zernike terms such as defocus and spherical aberration. Our work focuses instead on examples of $\Delta W$ which depend on $\xi$.

Before proceeding to the analysis of off-axis aberrations, we note that the spatial profile of $\psi$ is affected by the strength of the aberration $\Delta W$. For the purposes of quantifying this strength in later sections, it is useful to use $2\sigma_p$ and $2 \sigma$ as standard lengths for the pupil coordinate $u$ and the object location $\xi$, respectively. The reason for this choice is due to (1) the Gaussian pupil function strongly attenuating any signal at the pupil plane located at $|u| > 2\sigma_p$ and (2) the context of quantum-inspired resolution is most relevant for objects located at $|\xi| < 2 \sigma$; that is, for objects whose features are below twice the diffraction-limited PSF. This regime is henceforth referred to as the sub-Rayleigh regime.

\subsection{Off-axis aberrations}
From Eq.~(\ref{convreplace}), it is clear that the PSF of a shift variant system is determined by the aberration function $\Delta W(u,\xi)$. Although there is an infinitude of options; we restrict our analysis to the case where 
\begin{align}
    \Delta W(u,\xi) = W_{111} \left( \frac{\xi}{2 \sigma} \right) \left( \frac{u}{2\sigma_p} \right) + W_{220} \left( \frac{\xi}{2 \sigma} \right)^2 \left( \frac{u}{2\sigma_p} \right)^2, \label{aberrationfunction}
\end{align}
which is recognized as the superposition of the two low-order off-axis Seidel aberrations commonly called OAT and Petzval curvature \cite{wyant}. Notice that Eq.~(\ref{aberrationfunction}) is written in terms of ratios for $\xi$ and $u$ so that $W_{111}$ and $W_{220}$ can be interpreted as the maximal [when $\xi = 2\sigma$ and $u = 2\sigma_p$, as discussed in the paragraph following Eq.~(\ref{SVPSF})] optical path difference caused by OAT and Petzval curvature, respectively. By further defining $T$ and $P$ as
\begin{align}
    T \triangleq \frac{W_{111}}{\lambda \cdot (2\sigma)} \quad \text{and} \quad P \triangleq \frac{W_{220}}{\lambda \cdot (2\sigma)^2}, \label{strparams}
\end{align}
to be OAT and Petzval curvature \textit{strength parameters}, respectively (they measure rates, for $u = 2\sigma_p$, at which the phase difference caused by OAT and Petzval curvature, respectively, increases as a function of $\xi$), one finds that the intensity-normalized shift-variant PSF is given through Eq.~(\ref{SVPSF}) by
\begin{align}
    \psi(x,\xi;P,T) = \frac{1}{[2\pi g^2(\xi;P)]^{1/4}} \exp \left\{- \frac{[x - \xi(1 + 2\pi T \sigma)]^2}{4 g^2(\xi,P)} + \ii \Phi(x,\xi;P,T)\right\} , \label{PTPSF}
\end{align}
where
\begin{align}
    g(\xi;P) \triangleq \sigma \sqrt{1 + 4\pi^2 P^2 \xi^4}, \label{PSFwidth}
\end{align}
is the characteristic width of $\psi$ and 
\begin{align}
    \Phi(x,\xi;P,T) \triangleq -\frac{1}{2} \left\{ \tan^{-1}(2\pi P\xi^2) + \frac{\pi P \xi^2 [x - \xi(1+2\pi T \sigma)]^2}{ g^2 (\xi;P)} \right\}, \label{PSFphase}
\end{align}
is the phase of $\psi$. Equation~(\ref{PTPSF}) shows why the choice of Eq.~(\ref{aberrationfunction}) was used: OAT and Petzval curvature cause intuitive changes in the PSF's mean location and width, respectively. Therefore, much insight can be obtained from studying Eq.~(\ref{aberrationfunction}) since, while other off-axis aberrations may induce higher order effects on $\psi$, the effects of OAT and Petzval is expected to provide a useful leading-order description.

OAT and Petzval curvature, among other Seidel aberrations, are common in typical imaging systems and have well-known interpretations particularly in the language of geometrical (ray) optics. For a system with OAT, the plane wave emerging from the pupil associated with each object location $\xi$ is proportionally tilted (posses an additional linear phase). This often arises from unintentional tilts of the optics within the system and causes an incorrect magnification in the image. This magnification factor, according to Eq.~(\ref{PTPSF}) and Fig.~\ref{4fsystem}, is $(1+2\pi T \sigma)$; in the case of an object scene with two point sources, this magnification translates to an amplification of the separation in the image field. Petzval curvature, on the other hand, describes an imaging system in which the image is perfect when the intensity is measured along a curved surface. If the intensity is instead measured, as it usually is, along a flat surface, off-axis object points $\xi$ are mapped to a shift-variant PSF whose width increases with $|\xi|$. This width, according to Eqs.~(\ref{PTPSF}) and (\ref{PSFwidth}), this width is $g(\xi;P)$.

For the aberration-free case ($P = 0$ and $T = 0$), Eq.~(\ref{PTPSF}) reduces to the shift-invariant Gaussian PSF with width $\sigma$. Furthermore, it should be noted that the proceeding comparisons of Fisher information (FI) values between systems where the PSF is given by $\psi(x,\xi;P,T)$ and the aberration-free case $\psi(x,\xi;0,0)$ is fair, since $g(\xi;P) \ge \sigma$. In other words, the presence of OAT and Petzval curvature does not reduce the diffraction-limited spot size and therefore the improved FI is not attributed to the reduction of $\sigma$.

\subsection{Measurement schemes}
Although this work will primarily be concerned with incoherent imaging, a general framework is provided here for completeness and comprehension. To this end: the density matrix, which represents the cross-spectral density of the optical field at the image plane, for the case of two partially coherent point sources with intensity ratio $A$, a known centroid taken to be at the origin of the coordinate system, separated by a distance $s$ is given by
\begin{align}
    \rho(s;P,T) = \frac{|\psi_+ \rangle \langle \psi_+| + A|\psi_- \rangle \langle \psi_-| + \Gamma |\psi_+\rangle \langle \psi_-| + \Gamma^* |\psi_-\rangle \langle \psi_+|}{1 + A + 2 \text{Re}[\Gamma d(s;P,T)]}, \label{field}
\end{align}
which is expressed over the non-orthogonal, intensity-normalized basis $\{|\psi_+ \rangle, |\psi_-\rangle\}$. These basis kets are defined over position-space through Eq.~(\ref{convreplace}) as
\begin{align}
    |\psi_\pm\rangle &\triangleq \int_{-\infty}^\infty \psi_\pm(x;P,T) |x \rangle \, \ud x \triangleq \int_{-\infty}^\infty \psi\left(x, \pm \frac{s}{2} ;P,T\right) |x \rangle \, \ud x.
\end{align}
Furthermore,
\begin{align}
    d(s;P,T) \triangleq \langle \psi_- | \psi_+ \rangle = \int_{-\infty}^\infty \psi \left( x, \frac{s}{2} ; P, T\right) \psi^*\left(x, -\frac{s}{2}; P, T \right) \, \ud x
\end{align}
is the field-overlap integral between the two imaged PSFs and $\Gamma$ is a complex-valued coherence parameter whose values are restricted by the condition $|\Gamma| \le A$. The normalization factor in Eq.~(\ref{field}) indicates that the density matrix $\rho(s)$ uses the image-plane normalization scheme, in which the unit-trace condition on $\rho$ is ensured by the total number of \textit{received} photons at the image plane. This choice is in contrast to the object-plane normalization scheme; a detailed discussion of the two options is provided elsewhere.

\begin{figure}
    \centering
    \includegraphics[scale = 0.72]{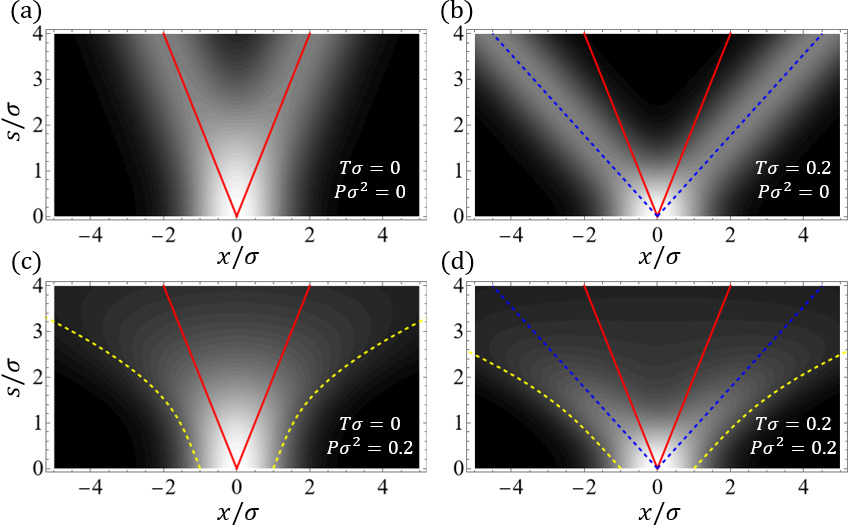}
    \caption{Density plots of $p_\text{I}(x,s;P,T)$ over $x/\sigma$ and $s/\sigma$ for various values of $T\sigma$ and $P\sigma^2$ in (a) - (d). In each, the solid red lines show the location of the two point sources. The dashed blue lines in (b) and (d) show the centers of intensity distribution from each point source for non-zero OAT. The dashed yellow lines in (c) and (d) show the outer envelope of $p_\text{I}$ for non-zero Petzval curvature.}
    \label{svpsf}
\end{figure}

The act of performing a measurement on the received photons is captured mathematically through computing the modulus-square coefficients over a corresponding set of projections of $\rho$. These coefficients are taken to be the probabilities of finding an image-plane photon in a certain mode (which may be continuous or discrete). In this work, we focus on comparing the measurement schemes of DI and SPADE. For DI, the set of projection modes is the continuous position basis $\{|x\rangle\}$; the probability density of finding a photon in the $x$ position of the image plane (conditioned on a detection event) is given by
\begin{align}
    p_\text{I}(x,s;P,T) &= \langle x| \rho(s;P,T) | x \rangle \nonumber\\
    &= \frac{|\psi_+(x;P,T)|^2 + |\psi_-(x;P,T)|^2 + 2 \text{Re}\left[\Gamma \psi_+(x;P,T) \psi_-(x;P,T) \right]}{1 + A + 2 \text{Re}[\Gamma d(s;P,T)]}. \label{probDI}
\end{align}
In Section~\ref{section:res}, we focus on the case where $\Gamma = 0$ and $A = 1$ (equally bright incoherent point sources). Figure~\ref{svpsf} is provided for the visualization of $p_\text{I}(x,s;P,T)$, which is a normalized version of the image-plane intensity distribution, for various levels of OAT and Petzval curvature. The aberration-free case shown in Fig.~\ref{svpsf}(a) shows the usual intensity pattern of two Gaussian PSFs converging as the separation $s$ vanishes. The effects of non-zero values of $T$ and $P$ are illustrated in Figs.~\ref{svpsf}(b) - (d): OAT causes the intensity distributions from the two point sources to converge at a different rate due to the magnification factor $(1+2\pi T \sigma)$ and Petzval curvature gives the intensity distributions a $s$-dependent width. Regarding OAT, it is clear from the comparison of Fig.~\ref{svpsf}(a) and (b) that non-zero values of $T$ allows for the two point sources' PSFs to be more easily distinguished for smaller values of $s/\sigma$. This effect, which will be detailed later, is responsible for larger values of CFI when using DI for imaging systems with OAT. Petzval curvature, on the other hand, does not have a similarly clear benefit in $s$-estimation whe compared to OAT. The resolution advantages offered by non-zero values of $P$ instead comes from the sensitivity of the width $p_\text{I}$ as $s$ approaches zero.

For SPADE, the projection is done over any set of discrete orthonormal modes. A natural choice is the $\sigma$-matched Hermite-Gauss (HG) basis $\{\phi_q(x)\}_{q=0}^\infty$, where the $q$-th mode is defined as
\begin{align}
    \phi_q(x) = \frac{1}{(2\pi \sigma^2)^{1/4}}\frac{1}{\sqrt{2^q q!}} H_q\left( \frac{x}{\sqrt{2} \sigma} \right) \exp \left( - \frac{x^2}{4\sigma^2} \right), \label{hgmodes}
\end{align}
and $H_q$ is the $q$-th physicist's Hermite polynomial. The probability of finding a photon (conditioned on a detection event) in the $q$-th mode is therefore the projection
\begin{align}
    p_\text{II}(q,s;P,T) &=  \langle \phi_q| \rho(s;P,T)|\phi_q \rangle  \nonumber\\
    &= \frac{|\langle \phi_q | \psi_+\rangle|^2 + |\langle \phi_q | \psi_-\rangle|^2 + 2 \text{Re}\left(\Gamma \langle \phi_q | \psi_+\rangle \langle \psi_- | \phi_q \rangle \right)}{1 + A + 2 \text{Re}[\Gamma d(s;P,T)]}. \label{probSPADE}
\end{align}
The probability densities given by Eqs.~(\ref{probDI}) and (\ref{probSPADE}) for DI and SPADE, respectively, are needed in the calculation of the CFI of either measurement scheme. For the case where only the separation $s$ is an unknown parameter, the corresponding CFIs for DI and SPADE are given by
\begin{align}
    F_\text{I}(s;P,T) &= \int_{-\infty}^\infty\frac{1}{p_\text{I}(x,s;P,T)} \left[ \frac{\partial}{\partial s}  p_\text{I}(x,s;P,T)  \right]^2 \, \ud x , \label{HDI}
\end{align}
and
\begin{align}
    F_\text{II}(s;P,T) &= \sum_{q = 0}^\infty \frac{1}{p_\text{II}(q,s;P,T)} \left[ \frac{\partial}{\partial s}  p_\text{II}(q,s;P,T) \right]^2, \label{HSPADE}
\end{align}
respectively. Although a full background on SPADE has been presented thus far, it is sufficient to consider a simplified version of SPADE known as binary SPADE (BSPADE) when one is interested in the sub-Rayleigh regime. In BSPADE, a photon arriving from the object plane is measured either in the $q = 0$ mode or in the combined $q>0$ mode. Intuition for this simplification comes from the fact that, as the object shrinks, most of the photons arriving at the image plane will project into the lowest order modes and the higher order modes contains progressively less information. In this case, the summation in Eq.~(\ref{HSPADE}) simplifies and can can be explicitly expressed as
\begin{align}
    F_{\text{II}}(s;P,T) \approx \frac{1}{p_\text{II}(0,s;P,T) - p_\text{II}^2(0,s;P,T)}  \left[ \frac{\partial}{\partial s}  p_\text{II}(0,s;P,T) \right]^2. \label{HBSPADE}
\end{align}
The quantities $F_\text{I}$ and $F_\text{II}$ are valuable in that, via estimation theory, their reciprocals provide lower bounds for the variance in the estimation of $s$ when the corresponding measurement scheme is used. In other words, if $\check{s}_\text{I}$ and $\check{s}_{\text{II}}$ are estimators for $s$ constructed from data acquired via DI or BSPADE measurements, respectively, then
\begin{align}
    \text{Var}(\hat{s}_\text{I}) \ge F_\text{I}^{-1} \quad \text{and} \quad \text{Var}(\hat{s}_\text{II}) \ge F_\text{II}^{-1}. \label{CFImeaning}
\end{align}
The CFI for DI and BSPADE measurements are compared for various values of $T$ and $P$ in Section~\ref{section:res}. There, it is shown that non-zero values of these aberration strength parameters can lead to larger values of $F_\text{I}$ and $F_\text{II}$.

\subsection{Quantum FI}
In addition to the CFI, which depends on the choice of measurement scheme, it is informative to consider the QFI, $Q_s$, which provides an upperbound to all possible CFI once the transformation between object and image planes is stipulated. In other words,
\begin{align}
    Q_s \ge F_\text{I} \quad \text{and} \quad Q_s \ge F_\text{II}. \label{QFIandCFI}
\end{align}
As is standard in deriving the QFI, one first finds the symmetric logarithm derivative (SLD), $L_s$, associated with the separation parameter $s$. The SLD is defined implicitly through 
\begin{align}
    \frac{\partial \rho(s;P,T)}{\partial s}  &= \frac{\rho(s;P,T) L_s + L_s \rho(s;P,T)}{2},
\end{align}
where $\rho$ is the image-plane density matrix given by Eq.~(\ref{field}); one should note that this prescription of $\rho$ corresponds to the image-plane normalization scheme detailed in []. With $L_s$ in hand, the QFI for $s$ is immediately obtained via
\begin{align}
    Q_s(s;P,T) &= \Tr \left[ \frac{\partial \rho(s;P,T)}{\partial s} L_s \right]. \label{QFI}
\end{align}
Details pertaining to the derivation of the QFI for imaging systems with off-axis aberrations are found in \textcolor{blue}{Supplement 1}. It should be emphasized that such calculations for the separation QFI for imaging systems in which the PSF is shift-variant require care and themselves constitute a novel result from which interesting discussions may arise. However, to keep the focus on well-known measurement schemes like DI and BSPADE, the QFI will primarily serve as a confirmation of CFI results through Eq.~(\ref{QFIandCFI}).

\section{Results and Discussion} \label{section:res}

\subsection{Results}
Although the framework developed in Section~\ref{theory} is general, we now specialize to the case of two equally bright incoherent point sources. This choice corresponds to the selection of $A = 1$ and $\Gamma = 0$ in Eqs.~(\ref{field}), (\ref{probDI}), and (\ref{probSPADE}). We are interested in calculating the CFI for DI and BSPADE for various values of $P$ and $T$ and comparing them to the case where the imaging system is aberration-free (and therefore shift-invariant), i.e., for $P = 0$ and $T = 0$.

\begin{figure}
    \centering
    \includegraphics[scale = 0.75]{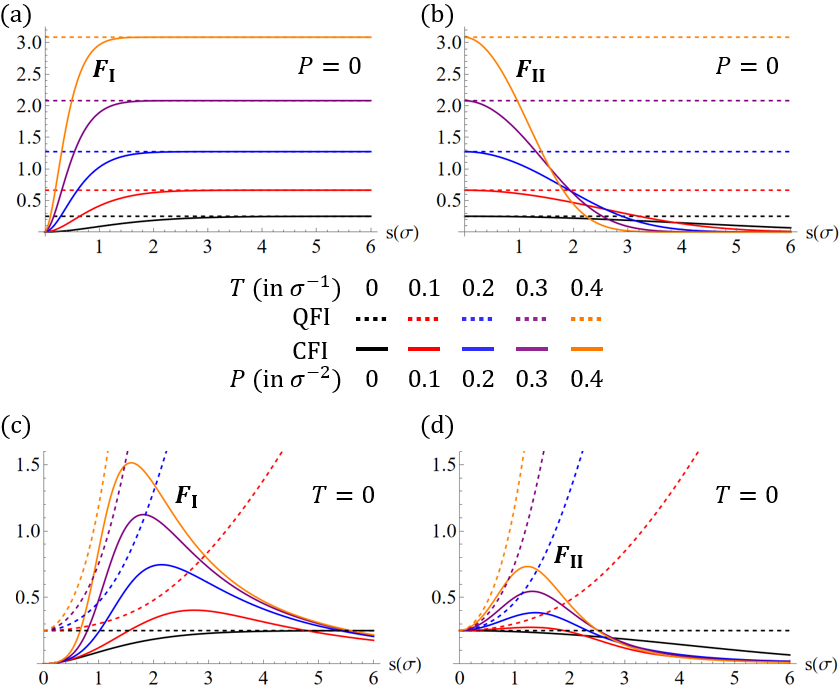}
    \caption{CFI (solid curves) associated with the separation $s$ for (a,c) DI and (b,d) BSPADE are shown for the case of (a,b) $P = 0$, varying $T$ and (c,d) $T = 0$, varying $P$. The QFI (dashed curves), $Q_s$, is also plotted for the corresponding values of $P$ and $T$. The black curves in each plot correspond to the aberration-free (shift-invariant) imaging system.}
    \label{PTplots1}
\end{figure}

In the following, the CFI for DI, given by Eq.~(\ref{HDI}), is calculated numerically. For BSPADE, it can be shown using Eqs.~(\ref{PTPSF}), (\ref{hgmodes}), and (\ref{probSPADE}) that
\begin{align}
    p_\text{II}(0,s;P,T) = \frac{1}{\sqrt{1 + P^2 \pi^2 (s/2)^4}}\exp \left\{ - \frac{(s/2)^2 (1 + 2\pi T \sigma)^2}{4 \sigma^2 [1 + P^2 \pi^2 (s/2)^4]} \right\},
\end{align}
which may be inserted into Eq.~(\ref{HBSPADE}) to obtain $F_\text{II}$. Comparisons of $F_\text{I}$ and $F_\text{II}$ for various values of $T$ and $P$ are displayed in Fig.~\ref{PTplots1}. When considering the $P = 0$ case (the imaging system only has OAT and no Petzval curvature), Fig.~\ref{PTplots1}(a) shows that increasing values of $T$ leads to greater $F_\text{I}$ for all values of $s/\sigma$: this global improvement in $F_\text{I}$ for non-zero values of $T$ is one of the main results of this analysis. For BSPADE, Fig.~\ref{PTplots1}(b) shows that a similar improvement in $F_\text{II}$ occurs when $T \neq 0$ in the sub-Rayleigh regime. Additionally, notice that $F_\text{II}$ saturates the QFI at exactly $s = 0$ for all values of $T$, which indicates that BSPADE continues to be an optimal measurement for $s$ even for shift-variant imagine systems. However, we re-emphasize that the main purpose of this analysis to to show that introducing off-axis aberrations in a system can vastly improve CFI even if one only considers a DI measurement scheme. The introduction of OAT with strength $T = 0.4 \sigma^{-1}$ leads to very high information (in comparison with the QFI value of $0.25$ in the aberration-free case) well into the sub-Rayleigh regime ($s < 2\sigma$). That is, even though $F_\text{I} = 0$ at exactly $s = 0$ always, one in principle can obtain any $F_\text{I}$ for non-zero $s/\sigma$ by increasing the strength of OAT even if $s/\sigma$ is small.

Figures~\ref{PTplots1}(c) and (d) consider the case where the imaging system contains Petzval curvature, but no OAT. It can be seen, in a behavior similar to that seen in Figs.~\ref{PTplots1}(a) and (b) that larger values of $P$ lead to larger values of $F_\text{I}$ and $F_\text{II}$ in the sub-Rayleigh regime, with the latter saturating the QFI near $s = 0$. Therefore, like OAT, Petzval curvature can improve the performance of an imaging system regarding two-point separation estimation. However, there are some notable differences between the behaviors of the CFI when the system contains OAT and when it contains Petzval curvature. First, increasing $P$ does not affect the value of the $F_\text{II}$ at exactly $s = 0$ ($F_\text{I} = 0$ regardless of $P$ and $T$). In other words, all the curves seen in Fig.~\ref{PTplots1}(d) coincide in the limit of vanishing separation. Second, there is an interval of $s$ within the sub-Rayleigh regime in which DI outperforms BSPADE. For example, comparing the $P = 0.4 \sigma^{-2}$ case for $F_\text{I}$ and $F_\text{II}$ in Figs.~\ref{PTplots1}(c) and (d), repsectively, it can be seen that the former reaches a peak of roughly $1.5$ compared to the latter's $0.75$. Finally, we note that Petzval curvature may lead to the formation of \textit{local} minima in $F_\text{I}$ and $F_\text{II}$, which indicate that the corresponding measurement schemes perform optimally neither at exactly $s = 0$ nor at $s \rightarrow \infty$, but rather at some intermediate value that, according to Figs.~\ref{PTplots1}(c) and (d), occur in the sub-Rayleigh regime.

\begin{figure}
    \centering
    \includegraphics[scale = 0.72]{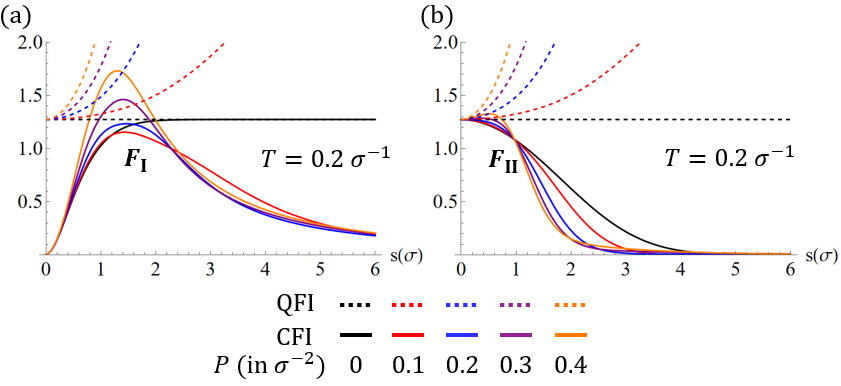}
    \caption{CFI (solid curves) associated with the separation $s$ for (a) DI and (b) BSPADE are shown for the case of $T = 0.2 \sigma^{-1}$, varying $P$. The QFI (dashed curves), $Q_s$, is also plotted for the corresponding values of $P$ and $T$.}
    \label{PTplots2}
\end{figure}

For completeness, CFI and QFI curves for the case where both OAT and Petzval curvature are present are shown in Fig.~\ref{PTplots2} for a fixed value of $T = 0.2 \sigma^{-1}$ and varying values of $P$. As was observed from Fig.~\ref{PTplots1}, larger values of $T$ raise the maximal attainable information while various values of $P$ lead to CFI curves that peak locally in the sub-Rayleigh regime [although such peaks for the BSPADE CFIs in Fig.~\ref{PTplots2}(b) are not apparent for smaller values of $P$ since the nonzero value of $T$ raises the CFIs at $s = 0$].

It is prudent to point out the behavior of the QFI curves in Figs.~\ref{PTplots1} and \ref{PTplots2} as well. When only OAT is present, the QFI remains constant over $s$, with its value given by
\begin{align}
    Q_s(s;0,T) = F_\text{II}(0;0,T) = \frac{(1 + 2\pi T \sigma)^2}{4\sigma^2}, \label{constantQFI}
\end{align}
which indicates that $Q_s(s;0,T)$ grows quadratically as a function of $T$. On the other hand, when Petzval curvature is present, the QFI increases from the value given by Eq.~(\ref{constantQFI}) at $s = 0$ to larger values as $s$ increases. This peculiar behavior, which is had not been seen for incoherent quantum-inspired superresolution studies of shift-invariant systems, is a result of the presence of the second term in the phase, $\Phi$, of the PSF given by Eq.~(\ref{PSFphase}). This quadratic (in $x$) phase contribution is due to the presence of Petzval curvature in $\Delta W$, as stipulated in Eq.~(\ref{aberrationfunction}); \textcolor{blue}{Supplement 1} provides more insight regarding the QFI's behavior.

\begin{figure}
    \centering
    \includegraphics[scale = 0.73]{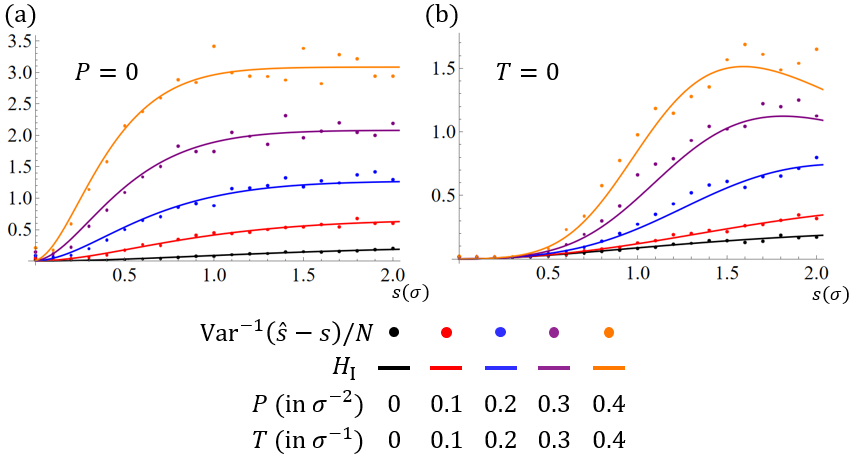}
    \caption{Comparisons of $F_\text{I}$ (solid curves) and $\text{Var}^{-1}(\hat{s}-s)/N$ from MLE simulations (discrete points) for the cases of (a) $P = 0$, varying $T$ and (b) $T = 0$, varying $P$. A mean photon number of $N = 2000$ was used for each iteration (each value of $s$) and the variance was calculated over $M = 500$ iterations.}
    \label{mlesimul}
\end{figure}

Maximum likelihood estimation (MLE) simulations for the separation $s$ were performed for DI to support the findings from Eq.~(\ref{HDI}) and Fig.~\ref{PTplots1}(a) and (c). For the $i$-th (out of $M$) iteration of the simulation, the number of photons arriving at the image plane, $N_i$, was chosen using Poisson statistics around an average photon number of $N$. That is, the probability mass function for $N_i$ is given by
\begin{align}
    p(N_i) = \frac{N^{N_i} \exp(-N)}{N_i!}.
\end{align}
Once $N_i$ is chosen, an equivalent number of photon positions in the image plane is chosen using Eq.~(\ref{probDI}) as the probability density function for $N_i$ independent events. Given these $N_i$ photon locations, the unbiased MLE estimator, denoted as $\hat{s}$, is used to find an estimate for the separation $s$. The variance of $\hat{s}$, whose average value is $s$, is then calculated over the ensemble of $M$ iterations. In order to compare the simulation results to $F_\text{I}$, one must divide the reciprocal of the aforementioned variance by $N$ to compute the per-photon CFI. This process is repeated for different values of $s, T,$ and $P$ and the results of these simulations are shown in Fig.~\ref{mlesimul} over the sub-Rayleigh regime $(s<2\sigma)$ for $N = 2000$ photons and $M = 500$ iterations. As is done in Figs.~\ref{PTplots1}(a) and (c), the cases of $P = 0$ and $T = 0$ were analyzed, respectively. It is evident that there is good agreement between the simulation results and $H_\text{I}$.

\subsection{Discussion}
The primary result of this work is that it is possible to improve upon the two-point resolution of an aberration-free imaging system by intentionally introducing known off-axis aberrations. Particularly, this is true even for DI measurement schemes as seen in Figs.~\ref{PTplots1} and \ref{PTplots2}. In fact, whether the improvement is global (over all $s$, as it is when OAT is present) or only over the sub-Rayleigh regime (when only Petzval curvature is present), the improvement can exceed the improvements offer by modal imaging techniques like BSPADE for aberration-free systems. In other words, if a sufficient amount of off-axis aberration can be provided, DI imaging schemes can lead to large $F_\text{I}$ even in the sub-Rayleigh regime. Since practical applications of two-point imaging deal with non-zero values of $s/\sigma$, increasing $T$ can effectively allow DI measurement schemes to attain the same information allowed by BSPADE in shift-invariant systems. However, this is not to say that BSPADE (and other modal imaging schemes) are fruitless; it is clear from Figs.~\ref{PTplots1} and \ref{PTplots2} that BSPADE benefits in a similar fashion to DI when off-axis aberrations are introduced. Importantly, at least for OAT and Petzval curvature, BSPADE still (as it did for the aberration-free case) yield a $F_\text{II}$ that saturates the QFI in the sub-Rayleigh regime.

The findings in this work are somewhat counter-intuitive as many traditional imaging systems start with substantial aberrations and a significant amount of effort is often needed to minimize such errors towards the goal of an aberration-free (diffraction-limited) system. However, at least in the context of resolving two incoherent and equally bright point sources where the separation is the only unknown parameter, it turns out that the presence of off-axis aberrations can actually improve performance. It may be of interest then, given the number of well-corrected imaging systems in existence, to understand how one may intentionally introduce OAT and Petzval curvature back into these imaging systems in order to take advantage of the larger DI or BSPADE CFI. Of course, it is possible to purposely (re)introduce aberrations by misaligning or adding/subtracting optical elements in the system. However, we should mention that despite their simple geometrical optics interpretations, OAT and Petzval curvature cannot be properly introduced by simply tilting or curving the image plane, respectively. Although these geometrical manipulations of the image surface create effective PSFs that are SV, they do not have the necessary dependence on $\xi$, the object plane coordinate, in order to bring about the results from a genuine $\Delta W$ given by Eq.~(\ref{aberrationfunction}).

\section{Concluding Remarks}
The analysis presented in this work inform on the CFI (for DI and BSPADE) and QFI regarding the separation estimation between two equally bright incoherent point sources when the imaging system includes off-axis aberrations. The resulting shift-variant nature of the system's field PSF gives rise to CFI and QFI that are novel and, importantly, \textit{greater} than their counterparts in the aberration-free case. Specifically, OAT and Petzval curvature, which constitute two of the lowest order off-axis Seidel aberrations, were shown to provide improvements to the CFI for both DI and BSPADE measurement schemes. A summary of their effects is provided as follows:
\begin{itemize}
    \item OAT [$\Delta W(\xi,u)$ is linear in both object ($\xi$) and pupil ($u$) locations] induces a magnification on the image field. In other words, two object points with separation $s$ are mapped to two PSFs with (larger) separation $(1+2\pi T \sigma)s$. This magnified separation provides the intuition for globally a larger CFI and QFI compared to the aberration-free case.
    \item Petzval curvature [$\Delta W(\xi,u)$ is quadratic in both object ($\xi$) and pupil ($u$) locations] induces an image field where the width of the PSF varies with object location. As two object points with separation $s$ approach each other $(s\rightarrow 0)$, the width of the two PSFs approaches the aberration-free case ($g \rightarrow \sigma$). This object-dependent width increases the sensitivity of the image field, which in turn leads to a larger CFI and QFI despite the fact that $g > \sigma$ for nonzero separation $s$.
\end{itemize}
QFI results were also developed and compared with the CFI. Once again, the details of the QFI derivation are shown in \textcolor{blue}{Supplement 1}.

To more fully appreciate the results of this work, it is valuable to contextualize them with the majority of the research done so far regarding quantum-inspired superresolution. The primary message in recent history is that novel modal imaging schemes, like BSPADE, can provide an advantage in resolution compared to traditional DI measurements. Such claims have been extended, and supported through theory and experiments, to more complicated object distributions (discrete or continuous). Furthermore, additional works have sought to optimize modal imaging. Examples include the development of practical adaptive imaging schemes that leverage advantages in both DI and BSPADE as well as a theoretical analysis on the effect of photon statistics \cite{Grace:20,ultimate2}. However, our present work demonstrates that the presence of off-axis aberrations, whose study have been largely ignored in the context of quantum-inspired superresolution, provides an improvement to both CFI and QFI that is relatively intuitive and whose mathematical treatment is straightforward.

The findings here are reminiscent of computational imaging techniques in which an imaging system is intentionally altered from the traditional aberration-free DI scheme in order to benefit from the known adjustments. We show that the introduction of OAT and Petzval curvature into imaging systems can improve the CFI in two-point separation estimation and therefore provide a link between the recent field of quantum-inspired superresolution with aspects of computational imaging. However, although the derivation presented here and in \textcolor{blue}{Supplement 1} for the CFI and QFI encompasses more than the case where the object scene consists of two equally bright incoherent point sources, realistic objects are much more complicated and require more care in their treatment. In addition to this concern, which is relevant in all of quantum-inspired superresolution analyses, realistic imaging systems have constraints regarding the strength of off-axis aberrations like OAT and Petzval curvature due to manufacturing and tolerancing limitations. Further studies are required to determine the benefits of introducing off-axis imaging systems to resolve general, realistic object scenes.
\newline \newline \noindent \textbf{Acknowledgements}. The author thanks S. A. Wadood, A. N. Vamivakas, and M. A. Alonso for useful discussions.
\newline \newline \noindent \textbf{Disclosures}. The author declares no conflicts of interest.
\newline \newline \noindent See \textcolor{blue}{Supplement 1} for supporting content.

\bibliography{sample}

\end{document}


\title{Off-axis Aberrations Improve the Resolution Limits of Incoherent Imaging: Supplement 1}

\author{Kevin Liang\authormark{1,2,*}}

\address{\authormark{1}Physics Department, Adelphi University, Garden City, NY 11530, USA\\
\authormark{2}The Institute of Optics, University of Rochester, Rochester NY 14627, USA\\
}

\email{\authormark{*}kliang@adelphi.edu} 



\begin{abstract*}
This document provides supplementary material for ``Off-axis Aberrations Improve the Resolution Limits of Incoherent Imaging". 
\end{abstract*}



\section{Point spread function for shift-variant systems}

In this section, a derivation via Fourier optics is provided for the field point spread function (PSF) of an imaging system that contains off-axis aberrations. Throughout the derivation, we will not be overly concerned with overall prefactors, as the PSF is ultimately normalized to ensure intensity-normalization at the image plane. Starting with a Dirac delta impulse at the object plane located at $\xi = \xi_0$, denoted $\delta(\xi - \xi_0)$ the propagation from the object plane to the pupil plane in a $4f$-imaging system is done through a Fourier transform. That is, the pupil plane field is given by
\begin{align}
    \delta(\xi - \xi_0) \rightarrow \int_{-\infty}^\infty \delta(\xi - \xi_0) \exp\left(- \ii k \frac{\xi u}{f} \right)\, \ud \xi
    &= \exp\left(- \ii k \frac{\xi_0 u}{f} \right).
\end{align}
Note that $k = 2\pi/\lambda$ is the wavenumber. Once at the pupil plane, there are two operations to consider. First, $U_p(u,\xi_0)$ encounters an aperture, which we model as a Gaussian with characteristic width $\sigma_p$. Second, $U_p(u,\xi_0)$ encounters a phase error (aberration) function $\Delta W(u,\xi_0)$. If $\Delta W$ depends only on the pupil-plane coordinates $u$, it is considered an on-axis aberration and typical examples of these include defocus, spherical aberration, and other aberrations often decomposed in terms of Zernike polynomials. If $\Delta W$ includes a dependence on $\xi_0$, the location of the original Dirac delta impulse, then it is considered an off-axis aberration; this is the case that is the focus of our work. To summarize, the transformation to $U_p(u,\xi_0)$ at the pupil plane is described as
\begin{align}
    \exp\left(- \ii k \frac{\xi_0 u}{f} \right) &\rightarrow \exp\left(- \ii k \frac{\xi_0 u}{f} \right)\exp\left( - \frac{u^2}{4 \sigma_p^2} \right)  \exp \left[ -\ii k \Delta W(u,\xi_0) \right].
\end{align}
The field is now inverse Fourier transformed to arrive at the image plane, where proper normalization gives us the definition of the PSF, $\psi$:
\begin{align}
    \psi(x,\xi_0) = U_0 \int_{-\infty}^\infty \exp\left( - \ii k \frac{\xi_0 u}{ f} \right)  \exp\left( - \frac{u^2}{4 \sigma_p^2} \right)  \exp \left[ -\ii k \Delta W(u,\xi_0) \right] \exp \left(\ii k  \frac{ux}{ f} \right)\, \ud u, \label{SVPSFsupp}
\end{align}
which is the definition seen in the main body, with $\xi_0 \rightarrow \xi$. Notice that, although an off-axis $\Delta W$ gives rise to shift-variance, the imaging system is still assumed to be linear. That is, if the object field, $U_o(\xi)$, is written a superposition of Dirac delta impulses:
\begin{align}
    U_o(\xi) = \int_{-\infty}^\infty U_0(\xi_0) \delta(\xi_0 - \xi) \, \ud \xi_0,
\end{align}
then 
\begin{align}
    U_i(x) = \int_{-\infty}^\infty U_o(\xi) \psi(x,\xi)\, \ud \xi.
\end{align}

\section{Quantum Fisher Information for shift-variant imaging systems} \label{suppsec1}

A framework for calculating the quantum Fisher Information (QFI) matrix associated with parameters to be measured from the image field of a shift-variant imaging system is provided in this section. These QFI calculations provide valuable upperbounds on the information associated with various quantities of interest (such as the separation between two point sources) when the system has off-axis aberrations. To keep the derivation general, we assume that the object scene is comprised of $N$ partially coherent point sources with point source locations $\{x_i\}_{i=1}^N$. Using the image-plane normalization (IN) framework in Ref.~\onlinecite{Liang2022rev}, the density matrix representing the object field is
\begin{align}
    \rho_\text{op} &= \sum_{i,j=1}^N \Gamma_{ij} |\xi_i \rangle \langle \xi_j|,
\end{align}
where $\Gamma$ is the object-plane mutual coherence matrix and $|\xi_i\rangle$ is the position ket at $\xi_i$, the location of the $i$-th point source. In the IN normalization framework, these kets may be harmlessly represented in position space as Dirac delta impulses:
\begin{align}
    \langle \xi|\xi_i \rangle = \delta(\xi-\xi_i) \label{DDimpulse}
\end{align}
The transition from the object to image planes can be captured by the blurring of each point source into the point spread function (PSF), $\psi$, of the imaging system. In many past analyses, imaging systems were treated as being shift-invariant and therefore the resulting image of each point source can be computed as the convolution of $\psi$ with the Dirac delta impulse, given by Eq.~(\ref{DDimpulse}) of each point source. However, for shift-variant systems, one must instead consider
\begin{align}
    |\xi_i \rangle \rightarrow |\psi_i\rangle,
\end{align}
where
\begin{align}
    |\psi_i\rangle &= \int_{-\infty}^\infty \psi(x,\xi_i) |x \rangle. \label{PSFSV}
\end{align}
Importantly, Eq.~(\ref{PSFSV}) indicates that the PSF $\psi(x,\xi_i)$ depends on both the image-plane position coordinate, $x$, and the point source's object plane location, $\xi_i$, in a manner that may not be expressible through their difference $x - \xi_i$ alone. We are primarily interested in the case where the imaging system contains off-axis tilt (OAT) and Petzval curvature. As seen in the main body, this leads to a normalized PSF given by
\begin{align}
    \psi(x,\xi;P,T) = \frac{1}{[2\pi g^2(\xi;P)]^{1/4}} \exp \left\{- \frac{[x - \xi(1 + 2\pi T \sigma)]^2}{4 g^2(\xi,P)} + \ii \Phi(x,\xi;P,T)\right\} , \label{PTPSF}
\end{align}
where
\begin{align}
    g(\xi;P) \triangleq \sigma \sqrt{1 + 4\pi^2 P^2 \xi^4}, \label{PSFwidth}
\end{align}
is the characteristic width of $\psi$ and 
\begin{align}
    \Phi(x,\xi;P,T) \triangleq -\frac{1}{2} \left\{ \tan^{-1}(2\pi P\xi^2) + \frac{\pi P \xi^2 [x - \xi(1+2\pi T \sigma)]^2}{ g^2 (\xi;P)} \right\}, \label{PSFphase}
\end{align}
is the phase of $\psi$. The values of $T$ and $P$ refer to the strength of OAT and Petzval, respectively (the case of $T = P = 0$ reduces the PSF to the shift-invariant, aberration free case). It's important to note that the \textit{width} of the PSF in Eq.~(\ref{PTPSF}) is such that $g \ge \sigma$, where $\sigma$ is the width of the aberration-free PSF; the equality is attained when $P = 0$ (no Petzval curvature). As is done in Refs.~\onlinecite{Liang2022rev} and \onlinecite{Bisketzi_2019}, it is convenient to re-express the PSF in terms by introducing the dimensionless position variable $\alpha = x/(2\sigma)$ so that the density matrix at the image plane can be written as
\begin{align}
    \rho_0 &= \mathcal{N}_0^{-1}\sum_{i,j=1}^N\Gamma_{ij} |\alpha_i \rangle \langle \alpha_j| \label{imageplane}
\end{align}
where
\begin{align}
    \langle \alpha |\alpha_i\rangle = \frac{1}{[\pi \bar{g}^2(\alpha_i;P)/2]^2} \exp \left\{ - \frac{[\alpha - \bar{h}(\alpha_i;T)]^2}{\bar{g}^2(\alpha_i;P)} \left[1 + \ii\bar{v}(\alpha_i;P) \right] - \frac{\ii}{2} \tan^{-1} [\bar{v}(\alpha_i;P)] \right\}. \label{nodimPSF}
\end{align}
Notice that the parameters to be estimated have transformed from $\{\xi_i\} \rightarrow \{\alpha_i\}$, where $\alpha_i = \xi_i/(2\sigma)$. Furthermore, we have introduced the dimensionless functions
\begin{align}
    \bar{h}(\alpha_i;T) &= \alpha_i(1 + 2\pi T \sigma),\label{barfunctionsh}\\
    \bar{g}(\alpha_i;P)&= \sqrt{1 + 64 \alpha_i^4\pi^2 P^2 \sigma^4},\label{barfunctionsg}\\
    \bar{v}(\alpha_i;P) &= 8 \pi \alpha_i^2 P \sigma^2. \label{barfunctionsv}
\end{align}
Note that, in the aberration-free case, $\bar{h}(\alpha_i;0) = \alpha_i$, $\bar{g}(\alpha_i;0) = 1$, and $\bar{v}(\alpha_i;0) = 0$. The normalization factor $\mathcal{N}_0$ in Eq.~(\ref{imageplane}) is given by
\begin{align}
    \mathcal{N}_0 &= \sum_{i,j=1}^N \Gamma_{ij} \int_{-\infty}^\infty \langle \alpha| \alpha_i\rangle \langle \alpha_j | \alpha \rangle \, \ud \alpha, \nonumber \\
    &= \sqrt{\frac{2 \bar{g}_i \bar{g}_j}{\bar{g}_i^2 (1 + \ii \bar{v}_j) + \bar{g}_j^2(1 - \ii \bar{v}_i)}} \exp \left[ -\frac{(\bar{h}_i -\bar{h}_j)^2 (\ii + \bar{v}_i)(1 + \ii \bar{v}_j)}{\bar{g}_i^2(\ii - \bar{v}_j) + \bar{g}_j^2(\ii + \bar{v}_i)} \right],
\end{align}
where we have introduced the shorthand $\bar{h}_i,\bar{g}_i,$ and $\bar{v}_i$ to for Eqs.~(\ref{barfunctionsh}), (\ref{barfunctionsg}), and (\ref{barfunctionsv}), respectively. It's worth emphasizing that the states in Eq.~(\ref{imageplane}) are not simply coherent states with fixed widths that are shifted to location $\alpha_i$. Instead, the centroid of the Gaussian and the width of the Gaussian are determined by the functions $\bar{h}$ and $\bar{g}$, respectively which \textit{depend} on object location $\alpha_i$. Such states are related to squeezed coherent states, where the centroid and widths of the Gaussian profile are fully correlated. The presence of such states in the expression for $\rho_0$ is a marked difference between systems that are shift-variant and the oft-studied aberration-free case.

It is also worth mentioning that the expression in Eq.~(\ref{nodimPSF}) can be used to represent the PSF of any system whose response to a point source located at $\alpha_i$ is a Gaussian centered at some location $\bar{h}(\alpha_i)$ with width $\bar{g}(\alpha_i)$. Although explicit functions for $\bar{h}$ and $\bar{g}$ are provided in the present analysis in Eqs.~(\ref{barfunctionsh}) and (\ref{barfunctionsg}) for the case of OAT and Petzval curvature, the following derivation for the QFI matrix can be straightforwardly adapted for any differentiable $\bar{h}$ and $\bar{g}$. Finally, the function $\bar{v}(\alpha_i;P)$ defined in Eq.~(\ref{barfunctionsv}) encapsulates the phasor portion of $|\alpha_i\rangle$. Although we keep our derivation general for an arbitrary mutual coherence matrix $\Gamma$, it is worth noting here that the final term within the curly braces in Eq.~(\ref{nodimPSF}) is irrelevant for QFI calculations when the object scene is incoherent. This is because the term is independent of $\alpha$ (a \textit{global} phase) and vanishes when Eq.~(\ref{imageplane}) is diagonal (incoherent object scene). However, the presence of $\bar{v}$ persists even in the incoherent case in the first term within the curly braces in Eq.~(\ref{nodimPSF}).

With the density matrix $\rho_0$ given in Eq.~(\ref{imageplane}), one can now proceed with QFI matrix calculations to obtain the precisions associated in the measurement of unknown parameters (a subset of $\{\alpha_i\}_{i=1}^N$). Given the complicated nature of $|\alpha_i\rangle$, we follow a general method of deriving the QFI matrix \cite{Liu2019,Liang2021,Bisketzi_2019}. A summary, and key differences/observations for the present case of a shift-variant imaging system, is provided as follows. One begins by identifying a basis that can be used to represent both $\rho_0$ as well as $\partial_j \rho_0$, where $\partial_j \triangleq \partial/\partial \alpha_j$ is a shorthand for parametric differentiation. Upon observation of Eq.~(\ref{imageplane}), it is clear that the union of $\{|\alpha\rangle_i\}$ and $\{\partial_i |\alpha_i\rangle\}$ is sufficient. With the sufficient basis identified, we now collect its elements in a $2N$-element vector
\begin{align}
    \vec{A} = \begin{bmatrix} |\alpha_1 \rangle & \cdots & |\alpha_N \rangle & \partial_1 |\alpha_1 \rangle & \cdots & \partial_N |\alpha_N \rangle \end{bmatrix}
\end{align}
so that we may express $\rho_0$ and $\partial_i \rho_0$ as
\begin{align}
    \rho_0 &= \vec{A}^\dagger \cdot \rho_{0,A} \cdot \vec{A},\\
    \partial_i \rho_0 &= \vec{A}^\dagger \cdot (\partial_i \rho_{0,A}) \cdot \vec{A}.
\end{align}
That is, $\rho_{0,A}$ and $\partial_i \rho_{0,A}$ are the coefficient matrices for the representation of $\rho_0$ and $\partial_i \rho_0$ in terms of the basis states collected in $\vec{A}$. Once this is done, one defines a $(2N)\times(2N)$ matrix $\Upsilon$, which is in turn defined in terms of submatrices:
\begin{align}
    \Upsilon \triangleq \begin{bmatrix} \Upsilon_{\alpha \alpha} & \Upsilon_{\alpha d} \\ \Upsilon_{d\alpha} & \Upsilon_{dd} \end{bmatrix}
\end{align}
The elements of these $N\times N$ Grammian submatrices are given by various inner products between basis states:
\begin{align}
    (\Upsilon_{\alpha \alpha})_{ij} &= \langle \alpha_i | \alpha_j \rangle  \label{Upsilonsubs1}\\
    (\Upsilon_{\alpha d})_{ij} &= \langle \alpha_i | \partial_j |\alpha_j \rangle \\
    (\Upsilon_{d \alpha})_{ij} &= (\Upsilon_{\alpha d}^\dagger)_{ij}\\
    (\Upsilon_{dd})_{ij} &= \langle \alpha_i| \partial_i^\dagger \partial_j | \alpha_j \rangle \label{Upsilonsubs},
\end{align}
where $\partial_i^\dagger$ indicates a derivative acting on the state to the \textit{left}. Although analytic formulas exist for these matrix elements, given the complicated and nested nature of the parameter $\alpha_i$ in Eq.~(\ref{nodimPSF}), their explicit expressions will not be stated here. The unruliness of these expressions is a marked difference between treating an shift-variant imaging system, where the PSF is given by Eq.~(\ref{nodimPSF}), and an aberration-free system. 

The elements of the QFI matrix, $\mathcal{Q}$, can then be shown to be given by
\begin{align}
    (\mathcal{Q}_0)_{ij} = 2\text{vecb}(\partial_i \rho_{0,A})^\dagger \cdot (\Upsilon^{-1} \odot \rho_{0,A} + \rho_{0,A}^*  \odot \Upsilon^{-1})^{-1} \cdot \text{vecb}(\partial_j \rho_{0,A}),
\end{align}
where $\text{vecb}(\cdot)$ is the block-column vectorization operator defined through the example on its action on $\Upsilon$ as 
\begin{align*}
    \text{vecb}(\Upsilon) = \begin{bmatrix} | \Upsilon_{\alpha \alpha} ) \\ | \Upsilon_{d \alpha} ) \\ | \Upsilon_{\alpha d} ) \\ | \Upsilon_{dd} ) \end{bmatrix},
\end{align*}
where $|\cdot )$ is the column vectorization operator on a matrix \cite{Liang2022rev}. In other words, $\text{vecb}(\cdot)$ takes a $(2N)\times(2N)$ matrix and stacks its four $N\times N$ submatrices column-wise before those submatrices themselves are vectorized to form a $4N^2$-element vector. Finally, $\odot$ is the Tracy-Singh block Kronecker product. By defining
\begin{align}
    \mathbb{S} &\triangleq \frac{q}{\sigma} \mathcal{N}_0^{-1} \left( \Upsilon_{\alpha \alpha}^{-1} \otimes \Gamma + \Gamma^* \otimes \Upsilon_{\alpha \alpha}^{-1} \right),\\
    \mathbb{B} &\triangleq \mathbb{I} \otimes \left( \Upsilon_{\alpha \alpha}^{-1} \Upsilon_{\alpha d} \right), \\
    \bar{\mathbb{B}} &\triangleq \left( \Upsilon_{\alpha \alpha}^{-1} \Upsilon_{\alpha d} \right) \otimes \mathbb{I},\\
    \mathbb{G} &\triangleq  \Upsilon_{dd} - \Upsilon_{d\alpha} \Upsilon_{\alpha \alpha}^{-1} \Upsilon_{\alpha d},
\end{align}
where $\otimes$ is the Kronecker product. The QFIM can in turn be expressed as
\begin{align}
    \mathcal{Q}_0 &= 2(\Xi + \Xi^\dagger + \Omega + \Omega^\text{T}),\label{QFIM4}
\end{align}
where $\Xi$ and $\Omega$ are matrices with elements
\begin{align}
    \Xi_{ij} &= \frac{(K_{0,i}^*|\mathbb{S}^{-1} |K_{0,j})}{2} + (K_{0,i}^*| \mathbb{S}^{-1} \mathbb{B} |Y_{0,j}) + (K_{0,i}^*|\mathbb{S}^{-1} \bar{\mathbb{B}}|Y_{0,j}^\dagger) + (Y_{0,i}^*| \mathbb{B}^\text{T} \mathbb{S}^{-1}\bar{\mathbb{B}}|Y_{0,j}^\dagger),\\
    \Omega_{ij} &= \left(Y_{0,i}^*\bigg|\mathbb{B}^\text{T}\mathbb{S}^{-1}\mathbb{B}+ \left( \mathcal{N}_0^{-1} \Gamma^*\right)^{-1} \otimes \mathbb{G}\bigg|Y_{0,j}\right),
\end{align}
respectively. Furthermore, 
\begin{align}
    |K_{0,i}) &\triangleq  \left[ \partial_i(\mathcal{N}_0^{-1}) |\Gamma) - \mathcal{N}_0^{-1} \alpha_i |F_i + F_i^\dagger) \right], \label{defK}\\
    |Y_{0,i}) &\triangleq  \mathcal{N}_0^{-1} |F_i),  \label{defH}
\end{align}
are $N^2$-dimensional column vectors and $F_i$ is a $N \times N$ matrix whose $i$-th row is the $i$-th row of $\Gamma$ and zero elsewhere. In other words,
\begin{align}
    (F_i)_{kl} &= \delta_{ik} \Gamma_{kl}.
\end{align}
Equation~(\ref{QFIM4}) is the QFI matrix corresponding to the set of unknown parameters $\{\alpha_i\}$ (or any subset of it), which are the individual locations of each point source. However, it is sometimes convenient to re-parameterize the problem in terms of the relative coordinates $\{s_i\}$, with
\begin{align*}
    s_i &= 2\sigma \begin{cases}
     \sum_{j=1}^N \alpha_j/N & i = 1,\\
    \alpha_i - \alpha_{i-1} & i > 1.
    \end{cases}
\end{align*}
Notice that $s_1$ is the centroid of the point sources and $s_{i>1}$ are successive differences between point source locations. The inverse Jacobian of this coordinate transformation is given by
\begin{align}
    [\mathbb{J}^{-1}]_{ij} &= \frac{\partial s_i}{\partial \alpha_j} = 2\sigma \left[ \frac{\delta_{i1}}{N} + \delta_{ij} - \delta_{(i-1)j}\right]. \label{invJacob}
\end{align}
The QFI matrix corresponding to the new parameters $\{s_i\}$, denoted $\mathcal{Q}'_0$, can be written, using the inverse of Eq.~(\ref{invJacob}) as
\begin{align}
    \mathcal{Q}'_0 = \mathbb{J}^\text{T} \mathcal{Q} \mathbb{J}. \label{sepMat}
\end{align}

Notice that the derivation provided above is for a general partially coherent object using the image-plane normalization scheme, and is provided for completeness. However, for the direct purposes of the present work, we are interested in the specific case where the object scene consists of two $(N=2)$ point sources that are equally bright and incoherent ($\Gamma_{ij} = \delta_{ij}/2$). By taking only the parameter $s_1$ to be unknown (the separation between the two point sources), the QFI matrix in Eq.~(\ref{sepMat}) becomes a single number corresponding to the QFI of estimating $s_1$: this value is the one used and plotted in the main body of this work and labeled $Q_s(s;P,T)$.

\section{Effect of the Petzval curvature phase term in PSF on QFI and CFI}
\begin{figure}
    \centering
    \includegraphics[scale = 0.72]{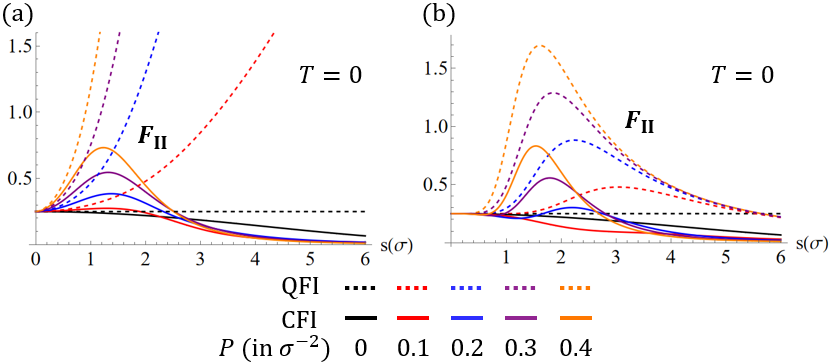}
    \caption{QFI (dashed) and CFI (solid) are shown for imaging systems where the PSF is given by $\psi$ (a) and $\bar{\psi}$ (b), which are given by Eqs.~(\ref{PSFSV}) and (\ref{PSFSV2}) respectively. There is no OAT ($T = 0$) and various values of $P$ correspond to the different colors.}
    \label{supp1}
\end{figure}

The motivation of this section is to give further insight into the divergent QFI for $P \neq 0$ seen in the main body. We compare the differences in QFI and CFI for imaging systems with field PSF given by Eq.~(\ref{PSFSV}) and
\begin{align}
    \bar{\psi}(x,\xi;P,T) = \frac{1}{[2\pi g^2(\xi;P)]^{1/4}} \exp \left\{- \frac{[x - \xi(1 + 2\pi T \sigma)]^2}{4 g^2(\xi,P)}\right\}. \label{PSFSV2}
\end{align}
It should be emphasized that Eq.~(\ref{PSFSV}) is the correct PSF to be used in QFI and CFI calculations when there is Petzval curvature; however, it is worthwhile to analyze Eq.~(\ref{PSFSV2}) to see the effect of the phase term $\Phi$ in Eq.~(\ref{PSFphase}).
Before presenting the QFI results, we note that Eq.~(\ref{PSFSV2}) alters the probability of photon detection at the lowest order Hermite-Gauss mode as $p_\text{II} \rightarrow \bar{p}_{\text{II}}$ where
\begin{align}
    \bar{p}_{\text{II}}(0,s;P,T) = \frac{\sqrt{1 + 4P^2 \pi^2 (s/2)^4}}{1 + 2P^2 \pi^2 (s/2)^4}\exp \left\{ - \frac{(s/2)^2 (1 + 2\pi T \sigma)^2}{4 \sigma^2 [1 + 2P^2 \pi^2 (s/2)^4]} \right\}.
\end{align}
This leads to an altered CFI for BSPADE via $F_\text{II} \rightarrow \bar{F}_\text{II}$, where
\begin{align}
    \bar{F}_{\text{II}}(s;P,T) \approx \frac{1}{\bar{p}_\text{II}(0,s;P,T) - \bar{p}_\text{II}^2(0,s;P,T)}  \left[ \frac{\partial}{\partial s}  \bar{p}_\text{II}(0,s;P,T) \right]^2.
\end{align}
Figure~\ref{supp1}(a), which also appears in the main body, shows the QFI and $F_\text{II}$ for the two-point separation in an imaging system where $\psi$ is the PSF, given by Eq.~(\ref{PSFSV}). As discussed within the main body, the QFI and $F_\text{II}$ coincide when $s \rightarrow 0$. However, the QFI diverges as the separation increases and is significantly different from $F_\text{II}$ even within the sub-Rayleigh regime of $s < 2\sigma$. On the other hand, when the imaging system is characterized by the PSF, $\bar{\psi}$, given by Eq.~(\ref{PSFSV2}), Fig.~\ref{supp1}(b) shows that the both the QFI and $\bar{F}_\text{II}$ are comparatively smaller. In particular, the QFI no longer diverges as $s$ increases. Therefore, the divergent behavior of the QFI for non-zero Petzval curvature is due to the presence of the phase, $\Phi$, within the PSF, $\psi$. 
Note that $\Phi$ does not affect the CFI for direct imaging (DI) since DI is an intensity-based measurement.

\bibliography{suppsample}